\documentclass[12pt]{iopart}

\pdfoutput=1
\expandafter\let\csname equation*\endcsname\relax
\expandafter\let\csname endequation*\endcsname\relax

\usepackage[english]{babel}
\usepackage{amsmath,
amssymb,amsfonts,latexsym}
\usepackage{graphicx}
\usepackage{mathrsfs} 
\usepackage{color}
\usepackage{subfigure}
\usepackage{afterpage}
\usepackage{booktabs}
\usepackage{blkarray}
\usepackage{cite}
\usepackage{tikz}
\usetikzlibrary{snakes}
\usetikzlibrary{calc}
\usepackage{bm}
\usepackage{braket}
\usepackage{nicefrac}

\definecolor{Blue}{rgb}{0.00, 0.00, 1.00}
\definecolor{Red}{rgb}{1.00, 0.00, 0.00}
\definecolor{labelkey}{cmyk}{.1,.7,0.5,0}

\definecolor{blue(pigment)}{rgb}{0.2, 0.2, 0.6}
\usepackage{mathrsfs}
\usepackage{times}
\usepackage[colorlinks,
citecolor=blue,linkcolor=blue,urlcolor=blue]{hyperref}
\hypersetup{linktocpage}
\usepackage{scalerel}
\usepackage{tensor}

\makeatletter
\def\@mkboth#1#2{}
\newlength\appendixwidth
\preto\appendix{\addtocontents{toc}{\protect\patchl@section}}
\newcommand{\patchl@section}{%
  \settowidth{\appendixwidth}{\textbf{Appendix }}%
  \addtolength{\appendixwidth}{1.5em}%
  \patchcmd{\l@section}{1.5em}{\appendixwidth}{}{\ddt}%
}
\makeatother

\def\eqref#1{(\ref{#1})}
\usepackage{cancel}

\newcommand{\be}{\begin{equation}}
\newcommand{\ee}{\end{equation}}
\newcommand{\bea}{\begin{eqnarray}}
\newcommand{\eea}{\end{eqnarray}}

\renewcommand*{\geq}{\geqslant}
\renewcommand*{\leq}{\leqslant}
\newcommand{\I}{\ensuremath{\mathbf{i}}}
\include{new_commands}

\newcommand{\arcth}{{\rm arcth}}
\newcommand{\dd}{{\rm d}}
\newcommand{\RR}{\mathbb{R}}
\newcommand{\ZZ}{\mathbb{Z}}
\newcommand{\Real}{\mathrm{Re\,}}
\newcommand{\Imag}{\mathrm{Im\,}}

\begin{document}

\title[]{Entanglement Hamiltonian during a domain wall melting in the free Fermi chain}
\author{Federico Rottoli$^{1}$, Stefano Scopa$^{1}$ and Pasquale Calabrese$^{1,2}$}
\address{$^1$ SISSA and INFN, via Bonomea 265, 34136 Trieste, Italy}
\address{$^2$ International  Centre  for  Theoretical  Physics  (ICTP),  I-34151,  Trieste,  Italy}
\date{\today}
\begin{abstract}
We study the unitary time evolution of the entanglement Hamiltonian of a free Fermi lattice gas in one dimension initially prepared in a domain wall configuration. To this aim, we exploit the recent development of quantum fluctuating hydrodynamics. 
Our findings for the entanglement Hamiltonian are based on the effective field theory description of the domain wall melting and are expected to exactly describe 
the Euler scaling limit of the lattice gas.
However, such field theoretical results can be recovered from high-precision numerical lattice calculations only when summing appropriately over all the hoppings up to 
distant sites.
\end{abstract}

\tableofcontents

\maketitle

\section{Introduction}\label{sec:intro}
In the last decade or so, we are witnessing a rapid development in the understanding of the in- and out-of-equilibrium properties of quantum integrable models in one spatial dimension \cite{cem-16,bbdv-22}.  
This current progress is also the result of a research that began much further in the past with the idea of reducing complicated many-body quantum models on the lattice to an effective field theory having the same properties in the low energy regime, thus unveiling some universal aspects of the quantum fluids \cite{Haldane1981a,Haldane1981b,Haldane1981c,GiamarchiBook}.
In this way, it has been possible to characterise the large-distance behaviour of correlation functions \cite{Haldane1981c,GiamarchiBook,KorepinBook,GogolinBook} and the entanglement \cite{Calabrese2004,Calabrese2009} of a large class of equilibrium homogeneous quantum systems. From this fertile ground, in recent years, results for the quantum correlations of non-homogeneous systems have been obtained by means of elegant Weyl transformations \cite{Allegra2016,Dubail2017,Brun2017,Brun2018,Scopa2020,Ruggiero2019,Bastianello2020,dsc-17}, thanks to which an inhomogeneous setting can be mapped to a flat space configuration under mild smoothness assumptions. 
Roughly in parallel, the dynamical problem was also studied with the goal of completing such asymptotic approach with a quantum hydrodynamic theory.
To date, thanks to the re-quantisation of the semi-classical evolution established by the generalised hydrodynamics \cite{Bertini2016,CastroAlvaredo2016}, it has been possible to obtain promising results for the large-scale dynamics of the entanglement entropy \cite{Ruggiero2020,Collura2020,Scopa2021a,Scopa2021b, Ruggiero2021} and to reproduce the numerical data obtained for the microscopic models with an impressive precision.
Although some aspects of this emerging quantum hydrodynamic theory still remain to be clarified \cite{f-17,Fagotti2020}, it is a matter of fact that it provides asymptotically exact results for the entanglement dynamics, extending our analytical capabilities far beyond the current computational and conceptual limits set by standard lattice formulations. 

This paper fits in this context with the goal of extending the hydrodynamic approach to the calculation of the entanglement (or modular) Hamiltonian $\hat{K}_A\propto -\log\hat\rho_A$, defined as the logarithm of the reduced density matrix $\hat\rho_A$ of the subsystem $A$. 
The entanglement Hamiltonian has attracted a large theoretical interest for multiple reasons. 
First, it encodes, in a single object, the complete information about the entanglement structure in the system under analysis,
e.g., since the pioneering work of H.~Li and F.~D.~M.~Haldane \cite{Li2008}, it is used as a fingerprint of the topological properties of the system under investigation.
 In fact, as clear from this early milestone, the topology of a state is neatly read off from the symmetry resolution of the eigenvalues (encoded in the eigenvectors, i.e., the Hamiltonian), while obtaining it from the entanglement entropies (i.e., the eigenvalues) would require tripartitions rather than bipartitions \cite{Levin2005,Kitaev2006}. At the physical level, this is a consequence of the general property that the entanglement Hamiltonian of a many-body subsystem encodes  information about {\it all its subsystems}, including all possible multipartitions and symmetry resolution. This is far more than the knowledge of the entropies and eigenvalues for a large compact subsystem.
Unfortunately, its exact calculation is typically very challenging and it often becomes out-of-reach for many interesting settings, including even some  non-interacting systems,
see e.g. \cite{Eisler2017,Eisler2020,Javerzat2021,Mintchev2020,Mintchev2021,DiGiulio2019,DiGiulio2020,ep-18,ep-13,abch-17,achp-18,Eisler2019,fr-19,it-87,ncc-09,Mintchev2022} for some exact  results both on the lattice and in the field theory limit. 
In this respect, a fundamental tool is the Bisognano-Wichmann theorem \cite{Bisognano:1975ih,Bisognano1976, Unruh1976,w-18}, which provides the exact entanglement Hamiltonian of an arbitrary relativistic 
quantum field theory in $d$ dimensions for a half-space cut  $A\equiv\left \{\vec{x}\in \RR^d:  x^1 > 0, x^0 = 0 \right \}$.
In conformal field theories (CFT), this results can be used to infer the modular Hamiltonian for simply-connected bipartitions mappable to the half-space by conformal 
transformations in two \cite{Cardy2016,wrl-18,wrl-16} and higher dimensions \cite{Hislop1982,Casini2009,Wong2009,Casini2011}.
In Ref.~\cite{Tonni2018}, the CFT result  has been extended to non-homogeneous systems exploiting the above mentioned Weyl transformation to flat space. \\
On the other hand, a parallel and intriguing line of research has been to construct artificially a microscopic Bisognano-Wichmann like entanglement Hamiltonian to capture the 
universal features of the reduced density matrices and to engineer, both numerically and experimentally, such lattice  Bisognano-Wichmann Hamiltonians 
to measure entanglement properties that otherwise would not be accessible \cite{Dalmonte:2017bzm}.
 Indeed, although the reduced density matrix $\hat\rho_A$ of the system is in principle accessible with quantum-state tomography  \cite{Pichler2016,Beverland2018}, its direct measurement is exponentially inefficient with the increase of the system size.  In this sense, a quantum simulation and spectroscopy of the  entanglement Hamiltonian $\hat{K}_A$ rather than of the wave function exploits the locality properties of the reduced density matrix to give an efficient alternative to quantum-state tomography for the probe of the entanglement properties of a quantum system, see e.g.~Ref.\cite{Dalmonte:2017bzm}.
This program has been pursued for many interesting physical protocols, both in and out of equilibrium \cite{Dalmonte:2017bzm,Giudici:2018izb,zcdr-20,ksz-21,kbe-21,zksz-22,mgdr-19,mvdc-22,zhh-19,pa-18,mgfd-20,zhhw-20}, opening up the possibility to overcome the entanglement barrier present in 
many physical situations. 
However, to date, predictions for the entanglement Hamiltonian in non-homogeneous non-equilibrium settings are still lacking.

In this work, we would like to fill this gap presenting an analysis of $\hat{K}_A$ for the prototypical setting of a free Fermi lattice gas initially prepared in a domain wall configuration $\ket{\Psi_0}\sim\ket{\bullet\dots\bullet\circ \dots \circ}$ and subsequently let to freely expand. This problem has been thoroughly studied in the literature, both with lattice techniques e.g. \cite{Antal1999, Karevski2002,Rigol2004,Platini2005,Platini2007, Hunyadi2004,Antal2008,Vidmar2015} and in the field theory regime e.g. \cite{Vicari2012,Alba2014,Allegra2016,Dubail2017,Messner2019,Scopa2021a,Scopa2021b}. In particular, it was one of the first  models for which a semi-classical hydrodynamics has been formulated in the modern language \cite{Antal1999,Antal2008}, the first non-equilibrium setting for which the entanglement dynamics has been computed via quantum hydrodynamics \cite{Dubail2017}, and one of the extremely rare cases for which a non-equilibrium CFT description of quantum fluctuations has been formulated \cite{Allegra2016}. 

{\it Outline.}~The paper is organised as follows. In Sec.~\ref{sec:model}, we introduce the model, the terminology and we briefly review the domain wall melting problem. Our main focus is on the hydrodynamic description in terms of the fermionic occupation function, for which we detail the semi-classical evolution in the phase space.  In Sec.~\ref{sec:quantum-hydro}, we re-quantise the problem employing an effective field theory at low energy.  We first revisit the equilibrium description of low-energy quantum fluctuations in terms of an inhomogeneous Luttinger liquid and we subsequently move to the melting dynamics, for which an effective field theory can be formulated as in Ref.~\cite{Allegra2016}. The core of this paper is Sec.~\ref{sec:ent-H}. In particular, in Sec.~\ref{sec:annulus}, we review the annulus method for the computation of the entanglement Hamiltonian and in Sec.~\ref{sec:entanglement} we use it to derive an exact asymptotic prediction for the R\'enyi entropy, first obtained in Ref.~\cite{Dubail2017} with the twist field approach. Section~\ref{sec:ent-H-calc} contains the derivation and the result of the entanglement Hamiltonian during the domain wall melting, which is the major result of this work. High-precision exact numeric calculations on the lattice model have been performed to test our findings, leading to an excellent agreement of the field theory prediction with the data. We report the numerical analysis in Sec.~\ref{sec:numerics}. Finally,  Sec.~\ref{sec:conclusion} contains our conclusions and a few outlooks.

\section{The model and the quench protocol}\label{sec:model}
\begin{figure}[t]
\centering
\includegraphics[width=0.85\textwidth]{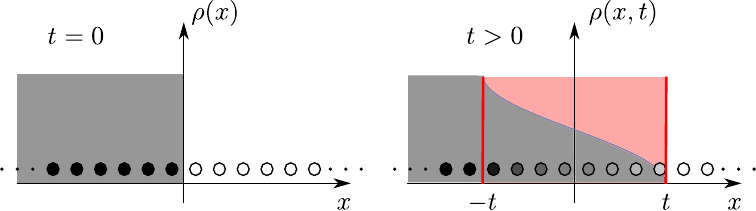}
\caption{Illustration of the domain wall setting. At $t=0$ the system is entirely filled on the l.h.s. and left empty on r.h.s.; the fermionic density is $\rho(x)=\Theta(-x)$. At $t>0$ the domain wall melts inside the light cone region $|x|\leq t$ ({\it light red region}) and the system develops a non-homogeneous density profile given by Eq.~\eqref{eqn:realdensity}.}\label{fig:DW-illustration}
\end{figure}

We consider a chain of free fermions particles with nearest-neighbour interactions loaded on an infinite one-dimensional lattice $i\in \ZZ$ and coupled to an external potential $V_i$, described by the Hamiltonian
\begin{equation}\label{H}
	\hat{H} = -\frac{1}{2} \sum_{i\in\ZZ} \left[\left(\hat{c}^\dagger_{i} \hat{c}_{i + 1} + \hat{c}^\dagger_{i+1} \hat{c}_{i}\right)+ V_i\ \hat{c}^\dagger_i\hat{c}_i\right].
\end{equation}
Here, $\hat{c}^\dagger_i$ (resp. $\hat{c}_i$) denotes the creation (resp. annihilation) operator of lattice spinless fermions satisfying canonical anticommutation relations $\{\hat{c}_i,\hat{c}_j^\dagger\}=\delta_{i,j}$. The system is initially prepared in the ground state of the Hamiltonian \eqref{H} with potential 
\be
V_i(t\leq 0)=\lim_{\Lambda\to\infty} \begin{cases} -\Lambda, \quad \text{if $i\leq 0$;} \\[4pt] \Lambda,\quad\text{otherwise,}\end{cases} \ee
which gives rise to the initial configuration
\be\label{initial-state}
\ket{\Psi_0} =\bigotimes_{i\leq 0} \ket{1}_i \; \bigotimes_{i> 0} \ket{0}_i,
\ee
where $\ket{\alpha=0,1}_i$ are the eigenstates of the number operator $\hat{c}^\dagger_i\hat{c}_i$ with eigenvalues $\alpha=0,1$. 
For times $t>0$, we set $V_i=0$ and we let the system to evolve unitarily with the hopping Hamiltonian \eqref{H}
\be\label{psi-t}
\ket{\Psi(t)}=e^{-\I t \hat{H}} \ket{\Psi_0}.
\ee
In other words, we investigate a fully-filled gas of hard-core particles initially confined on the left side of an infinite lattice and let to freely expand towards the right vacuum. 
The product state~\eqref{initial-state} is usually referred to as a {\it domain wall} and the quench dynamics \eqref{psi-t} as {\it domain wall melting}, 
in reference to the equivalent formulation in terms of the XX spin chain. 
Although the non-interacting nature of the underlying problem typically allows for exact lattice calculations, we rather consider its Euler hydrodynamic description where space-time scales $i,t\to\infty$ at fixed ratio $i/t$. Indeed, employing such a hydrodynamic description not only gives access to asymptotically exact results for conserved quantities \cite{Antal1999, Karevski2002,Rigol2004,Platini2005,Platini2007, Hunyadi2004,Antal2008,Vidmar2015} but it further allows to investigate several non-trivial properties of the model, including correlation functions \cite{Brun2017,Brun2018,Ruggiero2019,Scopa2020} and R\'enyi entropies \cite{Ruggiero2020,Bastianello2020,Collura2020,Scopa2021a,Scopa2021b}, which are currently not accessible by standard lattice techniques even for the free Fermi gas in such non-homogeneous and non-equilibrium settings. Hence, following this program,  the macrostate at $t=0$ is given by the fermionic occupation function \cite{Wigner1997, Hinarejos2012}
\be\label{Wigner}
n_0(x,k)=\begin{cases}
1, \quad\text{if $x\leq 0$ and $-\pi\leq k \leq \pi$};\\[3pt] 0, \quad \text{otherwise}
\end{cases}
\ee
as it reproduces, in the hydrodynamic limit, the initial domain wall state of Eq.~\eqref{initial-state} with left part of the system entirely filled with modes  $-\pi\leq k\leq \pi$ and right side left empty.
Notice that the lattice site $i$ is now replaced by a continuous variable $x\equiv i a \in \mathbb{R}$, where $a$ is the lattice spacing.
At times $t>0$, the evolution of $n_0(x,k)$ can be deduced from the trajectory of each mode $k$ that propagates independently with constant velocity $v(k)=\sin k$ from its initial position. This hydrodynamic picture leads to the macrostate
\be
n_t(x,k)\equiv n_0(x-t\sin k,k)=\begin{cases}
1, \quad\text{if  $k_F^-(x,t)\leq k \leq k^+_F(x,t)$};\\[3pt] 0, \quad \text{otherwise}
\end{cases}
\ee
with local Fermi points $k^\pm_F(x,t)$ given as solution of the zero-entropy hydrodynamic equation \cite{Doyon2017}
\be\label{zero-entropy-GHD}
\left(\partial_t +\sin k^\pm_F \partial_x\right) k^\pm_F=0.
\ee
More precisely, for a given time $t>0$ and position $0\leq x\leq t$, one finds the Fermi sea
\begin{equation}\label{eqn:realfermipoints}
\Gamma_t(x)\equiv\left[k_F^-(x,t);k_F^+(x,t)\right]=\left[\arcsin\frac{x}{t} ; \pi - \arcsin\frac{x}{t}\right]
\end{equation}
and an analogous treatment applies for $-t\leq x< 0$ exploiting the particle-hole symmetry of the problem. The fastest excitations of this setting are the modes $k=\pm\pi/2$ with velocity  $v(k =\pm \pi/2) = \pm 1$. These define the \emph{light-cone region} $|x| \leq t$ inside which correlations and entanglement spread during the quench dynamics and the particle density shows a non-homogeneous profile \cite{Antal1999, Antal2008}
\begin{equation}\label{eqn:realdensity}
	\rho(|x|\leq t) = \int_{\Gamma_t(x)} \frac{dk}{2\pi} = \frac{1}{\pi}\arccos\frac{x}{t}.
\end{equation}
Outside the light cone, i.e., for $x>t$ (resp. $x<-t$), the system keeps its initial configuration with fermionic density $\rho=0$ (resp. $\rho=1$). In Fig.~\ref{fig:DW-illustration}, we show an illustration of the domain wall state and of the melting dynamics considered in this work.
\section{Quantum hydrodynamic description}\label{sec:quantum-hydro}
The hydrodynamic theory outlined so far describes the semi-classical evolution in phase space of the free fermions but it does not account for the quantum fluctuations of the expanding gas. As such, it allows us to compute the semi-classical profiles of conserved quantities, but it is not sufficient for the study of entanglement. Since a microscopic derivation of the missing quantum effects is quite demanding, we look for an effective field theory description that is able to capture the relevant quantum processes in the low-energy regime.

\subsection{Equilibrium description of quantum fluctuations}\label{inhomogeneous-Lutt}
We first revisit the homogeneous gas at equilibrium, i.e. the ground state of the Hamiltonian \eqref{H} with  $V=0$. 
In this simple case, the correlation functions at large distances are effectively reproduced by expanding the lattice fermionic operators as \cite{GiamarchiBook}
\begin{equation}\label{c-exp-homo}
	\frac{\hat{c}_i}{\sqrt{a}} \sim e^{-\I k_F x} \psi_R(x) + e^{\I k_F x} \psi_L(x),
\end{equation}
where $\psi_{L}$ ($\psi_R$) is the left- (right-) moving chiral component of a massless Dirac fermion, whose action in imaginary time $\tau$ reads ($z\equiv x +\I \tau$)
\begin{equation}
	S = \frac{1}{2\pi}\int \left [\psi_R^\dagger \ \partial_{\bar{z}} \ \psi_R + \psi_L^\dagger \ \partial_{z} \ \psi_L \right ] \dd^2z.
\end{equation}
In the inhomogeneous case $V\neq 0$, the trapped gas is characterised by a spatially-dependent Fermi velocity $v_F(x)=\sin k_F(x)$. As a consequence, the effective field theory description in terms of a massless Dirac fermion requires a non-flat metric with line element \cite{Allegra2016,Dubail2017}
\begin{equation}\label{eqn:curvedstaticmetric}
	\dd s^2 = \dd x^2 + v_F^2(x) \dd \tau^2.
\end{equation}
It is then useful to find a set of \emph{isothermal coordinates} $z, \bar{z}$ in terms of which the metric is flat up to a Weyl factor, $\dd s^2 = e^{2\sigma(z, \bar{z})} \dd z \dd\bar{z}$ \cite{Dubail2017}. For the metric in Eq.~\eqref{eqn:curvedstaticmetric}, a simple choice is given by
\begin{equation}
	z(x, \tau) = \int^x \frac{\dd u}{v_F(u)} +\I \tau = \tilde{x} + \I \tau
\end{equation}
and the Weyl factor is equal to the Fermi velocity $e^{\sigma(x)} = v_F(x)$. Indeed, using isothermal coordinates, the action of the Dirac fermion in curved space takes the simple form
\begin{equation}\label{eqn:curvedDiracaction}
	S = \frac{1}{2\pi} \int e^{\sigma(z, \bar{z})} \left [ \psi_R^\dagger \ \partial_{\bar{z}}\  \psi_R + \psi_L^\dagger \ \partial_{z} \ \psi_L \right ] \dd^2z.
\end{equation}

Notice that under a Weyl transformation $e^{2 \sigma(z, \bar{z})} \dd z \dd\bar{z} \to \dd z \dd\bar{z}$ to flat space, a primary field $\phi$ of scaling dimension $\Delta$ transforms as
\begin{equation}
	\phi(z, \bar{z}) \to e^{- \Delta \sigma(z, \bar{z})} \phi(z, \bar{z}).
\end{equation}
It is then possible to obtain the asymptotic behaviour of the correlations of primary fields in curved space from the knowledge of those computed in a flat geometry, see e.g. Refs.~\cite{Brun2017,Brun2018,Ruggiero2019,Scopa2020}. Another important consequence of the position-dependent Fermi momentum is the phase appearing in the expansion of the lattice fermions in terms of $\psi_{L,R}$ (cf. Eq.~\eqref{c-exp-homo})
\begin{equation}\label{eqn:latticefermion}
	\frac{\hat{c}_i}{\sqrt{a}} \sim e^{-\I \varphi^+(x)} \psi_R(x) + e^{-\I\varphi^-(x)} \psi_L(x)
\end{equation}
where $\varphi^\pm(x)$ is defined through the position-dependent differential phase
\begin{equation}\label{eqn:staticdiffphase}
	\dd \varphi^\pm (x) = \pm k_F(x) \dd x .
\end{equation}

\subsection{Effective field theory in the arctic circle}\label{sec:arctic}
\begin{figure}[t]
\centering
\includegraphics[width=0.45\textwidth]{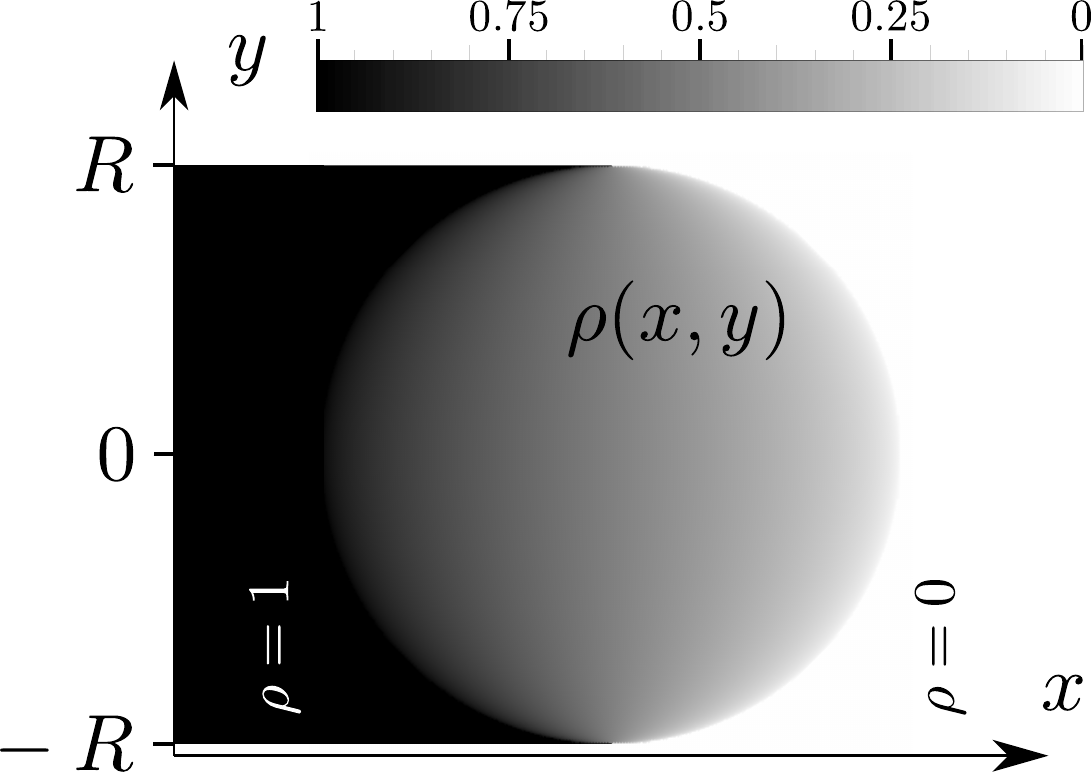}
\caption{Illustration of the fermionic density of Eq.~\eqref{dens-arctic} in the euclidean strip $\mathscr{S}$: $\rho(x,y)$ takes non-trivial values only inside the disk $x^2+y^2\leq R^2$, outside which it matches the boundary conditions imposed by the initial domain wall configuration.}\label{fig:arctic}
\end{figure}
We are now looking for an effective field theory which captures the quantum fluctuations of the domain wall quench problem discussed in Sec.~\ref{sec:model}. 
Following Refs.~\cite{cc-05,cc-06,Allegra2016,Dubail2017}, we study the non-equilibrium dynamics in imaginary time $y\equiv \I t\in [-R,R]$, introducing a finite width $R$ in the imaginary time direction. Doing so, the original quench problem is mapped in the euclidean strip $\mathscr{S}=\mathbb{R}\times [-R,R]\subset \mathbb{R}^2$, with boundary conditions set to reproduce the initial domain wall configuration. In this geometry, one can show that the fermionic density has a non-trivial profile only inside the disk $x^2 + y^2 \leq R^2$, typically referred to as {\it arctic circle} \cite{Allegra2016}
\begin{equation}\label{dens-arctic}
	\rho(x, y) = \begin{cases}
			1,                                                             &x < - \sqrt{R^2 - y^2}; \\
			\frac{1}{\pi} \arccos \left(\frac{x}{\sqrt{R^2 - y^2}}\right), &|x| \leq \sqrt{R^2 - y^2}; \\
			0,                                                             &x > \sqrt{R^2 - y^2},
		\end{cases}
\end{equation}
and outside which it matches the boundary conditions imposed by the initial domain wall, see Fig.~\ref{fig:arctic} for an illustration. The real time evolution is recovered by first performing an analytic continuation  to real time $y \to \I t$ and then by taking the limit $R \to 0$ \cite{cc-05,cc-06}. 
With this prescription, Eq.~\eqref{dens-arctic} reduces to the fermionic density in Eq.~\eqref{eqn:realdensity}. 
Inside the strip $\mathscr{S}$, the Fermi points \eqref{eqn:realfermipoints} become \cite{Allegra2016}
\begin{equation}\label{Fermi-arctic}\begin{matrix}
	k_F^+(x, y) = z(x, y);\\[5pt]
	k_F^-(x, y) = - \bar{z}(x, y),
\end{matrix}\end{equation}
where we introduced the coordinate
\begin{equation}\label{eqn:DWisothermal}
	z(x, y) = \arccos\left ( \frac{x}{\sqrt{R^2-y^2}} \right ) - \I \ \arcth\frac{y}{R}.
\end{equation}
Indeed, performing the continuation to real time $- \I \ \arcth\frac{y}{R}= \arctan\frac{t}{R} \to  \frac{\pi}{2}$, thus recovering the real time Fermi points of Eq.~\eqref{eqn:realfermipoints}.

We are now ready to investigate the quantum fluctuations around the Fermi points \eqref{Fermi-arctic}. 
Quantum fluctuations take place only inside the arctic circle, which is nothing but the light-cone region $|x|\leq t$ in imaginary time. 
Analogously to the case at equilibrium, the effective field theory for the quench problem is the one of a massless Dirac fermion in curved spacetime, with action given in Eq.~\eqref{eqn:curvedDiracaction}. For this problem, both the metric and the isothermal coordinates were found in Ref.~\cite{Allegra2016}. In particular, the line element is
\begin{equation}\label{line-element-arctic}
	\dd s^2 = \dd x^2 + \frac{2xy}{R^2 - y^2} \dd x \ \dd y + \frac{R^2 - x^2}{R^2 - y^2} \ \dd y^2,
\end{equation}
for which the isothermal coordinates are given by $(x,y)\to (z,\bar{z})$, with $z(x,y)$ defined in Eq.~\eqref{eqn:DWisothermal}. 
In real time, this set of coordinates corresponds  to the parametrisation of the right and left movers with their Fermi points. One can verify that, in terms of these coordinates,
the metric \eqref{line-element-arctic} becomes proportional to the flat one
\begin{equation}
	\dd s^2 = e^{2\sigma(x,y)} \dd z \ \dd\bar{z}
\end{equation}
with Weyl factor
\begin{equation}
	e^{\sigma(x,y)} = \sqrt{R^2 - y^2 - x^2 }.
\end{equation}
For later purposes, we mention that the time evolved fermionic operator $\hat{c}_i(t)$ picks up a space-time dependent semi-classical phase $\varphi_\pm(x,t)$ (similarly to Eq.~\eqref{eqn:latticefermion}), defined through the differential \cite{Dubail2017,Ruggiero2019}
\be
\dd \varphi_\pm(x,t) =k_F^\pm(x,t) \ \dd x - \varepsilon(k_F^\pm(x,t),x) \ \dd t.
\ee

We  remark that the knowledge of such non-equilibrium effective field theory in a curved space-time and, particularly, its reduction to a conformally flat one is a highly non-trivial result. To our best knowledge, the domain wall melting and the dynamics of the Tonks-Girardeau gas in a time-dependent harmonic trap (see Ref.~\cite{Lewis1969,Leach1982,Kagan1996,Minguzzi2005,Scopa2017,Scopa2018,Ruggiero2019}) are the only two out-of-equilibrium inhomogeneous cases for which an isothermal set of coordinates has been found \cite{Allegra2016,Ruggiero2019}. In more general situations, one can construct an effective field theory for the initial non-homogeneous state following the procedure outlined in Sec.~\ref{inhomogeneous-Lutt} and determine the dynamics of the quantum fluctuations using quantum generalised hydrodynamics, see e.g. \cite{Ruggiero2020,Collura2020,Scopa2021a,Scopa2021b}.

We also mention that this description has been used in Ref.~\cite{Dubail2017} for the calculation of the entanglement entropy with the twist field approach \cite{Calabrese2004,Calabrese2009, Cardy2008}. In the following, we will consider instead the  annulus method \cite{Cardy2016, Tonni2018} that will allow us to derive exact asymptotic results for the entanglement Hamiltonian, alongside recovering the known results for the entanglement entropies.

\section{Calculation of the entanglement Hamiltonian}\label{sec:ent-H}
We now move to the analysis of the entanglement spreading during the melting dynamics. As anticipated, although this problem has been already fully characterised in Ref.~\cite{Dubail2017}, we shall present in the following an alternative derivation  that also allows us to derive an asymptotically exact prediction for the entanglement Hamiltonian. 

We begin with a brief recap of some proprieties of entanglement measures.
Given a bipartition of the system $A\cup B$ at fixed time $t$, the R\'enyi entropy of index $n\in \mathbb{N}$ is defined as
\begin{equation}\label{Renyi-entropy}
	S_A^{(n)} = \frac{1}{1-n} \log \left [ \Tr \hat\rho_A^n \right ]
\end{equation}
where  $\hat\rho_A=\Tr_B\hat\rho$ is the reduced density matrix in $A$ obtained after tracing out the degrees of freedom of subsystem $B$ from the total density matrix $\hat\rho=\ket{\Psi(t)}\bra{\Psi(t)}$. 
Taking the \emph{replica limit} $n \to 1$, the R\'enyi entropy reduces to the von Neumann entropy (or entanglement entropy)
\begin{equation}
	S_A\equiv \lim_{n\to1} S_A^{(n)} = - \Tr \left [ \hat\rho_A \log \hat\rho_A \right ].
\end{equation}
Albeit the R\'enyi entropies proved to give a quite deep information on the quantum correlations of the many-body state and become recently accessible in cold-atom and ion-trap experiments \cite{Islam2015, Kaufman2016,Elben2018,Brydges2019,Elben2020,Neven2021}, 
a more comprehensive characterisation of the quantum state is provided by the entanglement (or modular) Hamiltonian $\hat{K}_A$
\begin{equation}
	\hat{K}_A=- \frac{1}{2\pi} \log \hat{\rho}_A.
\end{equation}

A fundamental result about the structure of the entanglement Hamiltonian of an arbitrary relativistic quantum field theory in $d$ dimensions is the Bisognano-Wichmann theorem \cite{Bisognano:1975ih,Bisognano1976, Unruh1976,w-18}. 
The latter states that the entanglement Hamiltonian of the ground state for a half-space cut  $A\equiv\left \{\vec{x}\in \RR^d:  x^1 > 0, x^0 = 0 \right \}$ corresponds to the 
generator of Lorentzian boosts (or, in imaginary time, of euclidean rotations):
\begin{equation}\label{eqn:BisWich}
	\hat{K}_A = \int_\text{A} x^1 T_{00}(x) \ \dd^{d-1} x
\end{equation}
where $T_{00}$ is the $00$ component of the stress-energy tensor. 
In $2d$ conformal field theories, it is possible to use the Bisognano-Wichmann result in Eq.~\eqref{eqn:BisWich} to obtain the entanglement Hamiltonian for all subsystem geometries that are are topologically equivalent to a cylinder by a simple conformal map \cite{Cardy2016}. 
Performing this map, the entanglement Hamiltonian can be written as the weighted integral of  $T_{00}$ \cite{Cardy2016} 
\begin{equation}\label{eqn:localinversetemp}
	\hat{K}_A = \int_A \frac{T_{00}(x)}{f^\prime(x)} \dd x\equiv \int_A \beta(x) T_{00}(x) \dd x
\end{equation}
where $f(x)$ is the projection at fixed (real) time of the analytic function that conformally maps the subsystem geometry into an annulus  (see Sec.~\ref{sec:annulus} below) while $\beta(x)\equiv 1/f^\prime(x)$ plays the role of a local inverse temperature, in analogy with the thermal ensemble for which $\hat{K}_\text{thermal} = \frac{\beta}{2\pi} \int T_{00}(x) \dd x $. 

\subsection{The annulus method}\label{sec:annulus}
In Refs.~\cite{Cardy2016}, it has been shown that it is possible to study the entanglement of $2d$ boundary conformal field theories by mapping the original geometry into an \emph{annulus}. Furthermore, with the help of the Weyl transformation discussed in Sec.~\ref{sec:quantum-hydro}, this method has been applied also to non-homogeneous systems \cite{Tonni2018}. In the following, we wish to extend this procedure to non-equilibrium inhomogeneous settings, building on some preliminary considerations put forward in Refs.~\cite{Cardy2016, Tonni2018}. We shall briefly review the annulus method before considering the specific case of a domain wall melting.

Let us consider a $2d$ boundary conformal field theory defined on a geometry $(x,y)\in\mathscr{G}\subset \mathbb{R}^2$. We further consider a cutting point $(x_0,y_0)$ and we investigate the entanglement in the subsystem $A$ between the position $x_0$ and the boundary of $\mathscr{G}$. Following Ref.~\cite{Cardy2016, Callan1994, Holzhey1994}, we introduce a UV regularization of the theory by removing  a small circle of radius $2\epsilon$ around $(x_0,y_0)$ and we map the resulting geometry  (which has the topology of a finite cylinder) into an annulus with a conformal transformation $z(x,y)\to w(z)$. The latter may be viewed as a rectangle of width $\mathcal{W}_A$ in the real direction and length $2\pi$ in the imaginary one with the identification $\Imag w +2\pi \equiv \Imag w$. 
Once the mapping from the original geometry $\mathscr{G}$ to the annulus is performed, the entanglement Hamiltonian is obtained as the generator of translations in the imaginary direction \cite{Cardy2016, Tonni2018}
\begin{equation}\label{K-annulus}
	\hat{K}_A  \equiv \int_{v=\text{cst}} T_{vv} \ \dd u=  \int_{w(A)} T_R(w) \ \dd w + \int_{\bar{w}(A)} {T}_L(\bar{w}) \ \dd \bar{w},
\end{equation}
where $w = u +\I v$ are the coordinates of the annulus, $T_L$ ($T_R$) are the chiral components of $T$ and $w(A)$ is the image of the subsystem $A$. In other words, it is sufficient to find an inverse map from the annulus  back to $\mathscr{G}$  (for us, the arctic circle) to obtain the entanglement Hamiltonian using Eq.~\eqref{K-annulus}. 
Since the R\'enyi entropy \eqref{Renyi-entropy} in terms of $\hat{K}_A$ is
\begin{equation}
S^{(n)} = \frac{1}{1-n} \log {\rm Tr} e^{-2\pi n \hat{K}_A},
\end{equation}
it is just related to the width $\mathcal{W}_A$ as \cite{Cardy2016, Tonni2018}
\begin{equation}\label{Renyi-cft}
	S^{(n)} = \frac{c}{12} \frac{n+1}{n} \mathcal{W}_A,
\end{equation}
where $c$ is the central charge of the conformal field theory, $c=1$ for the free Fermi gas.
Notice that the aforementioned UV regularisation enters in $S^{(n)}$ only through ${\cal W}_A$. 
The exact expression of the UV cutoff appearing in Eq.~\eqref{Renyi-cft} cannot be found within the field theory framework but requires exact lattice calculations, e.g. based on the Fisher-Hartwig conjecture, see \cite{Jin2004, Calabrese2010} and Sec.~\ref{sec:entanglement} below. In particular, in homogeneous systems such non-universal term would simply amount to an additive constant while, in non-homogeneous cases,  it carries a non-trivial spatial dependence  \cite{Dubail2017}.

We now  apply the annulus method to the domain wall melting problem. 
We set the imaginary time to a value $y_0$ such that $- R \leq y_0 \leq R$ and we study the entanglement of a bipartition with cut at position  $x_0$. 
As seen in Sec.~\ref{sec:arctic},  quantum fluctuations are present only inside the arctic circle. 
Therefore, the subsystem $A$ of interest is given by the intersection of the (regularised) right subchain $(x_0 + 2\epsilon, +\infty)$ with the arctic circle, i.e.,
\begin{equation}\label{A}
	A = \left \{ (x, y) \; :  x \in \left (x_0 + 2\epsilon, \sqrt{R^2 - y_0^2} \right ], y = y_0  \right \}.
\end{equation}

\begin{figure}
	\centering
	\includegraphics[width=\linewidth]{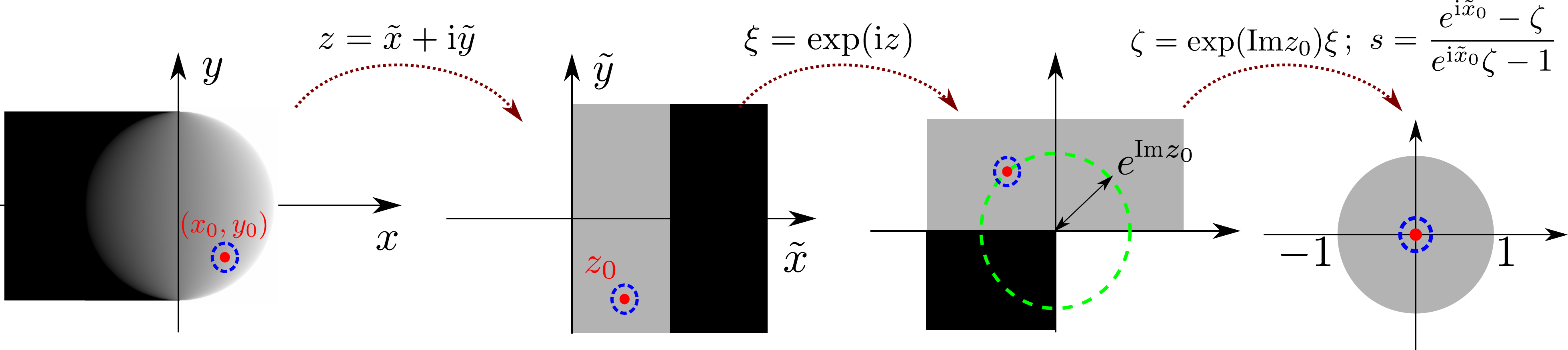}
	\caption{Illustration of the conformal mapping from the arctic circle to the annulus.}\label{fig:mapping}
\end{figure}

The conformal transformation to the annulus is shown in Fig.~\ref{fig:mapping}. The first step consists in a Weyl transformation to isothermal coordinates $(x,y)\to (z,\bar{z})$, with $z(x,y)\equiv \tilde{x}+\I \tilde{y}$ given in Eq.~\eqref{eqn:DWisothermal}, which maps the arctic circle into the flat strip $(\tilde{x},\tilde{y}) \in [0, \pi]\times \RR$. In particular, the entangling point $(x_0,y_0)$ is mapped to $z_0 = z(x_0, y_0) = \tilde{x}_0 + \I \tilde{y}_0$, while the boundary point $(\sqrt{R^2-y_0^2},y_0)$ on the arctic circle maps to
\begin{equation}\label{bound-arctic}
	z \left(x\equiv\sqrt{R^2 - y_0^2}, y\equiv y_0\right) = \I \tilde{y}_0.
\end{equation}
Under the Weyl transformation, the UV regularization changes as
\begin{equation}
	\tilde{x}(x + 2\epsilon, y) \approx \tilde{x}_0  - \frac{2\epsilon}{\sqrt{R^2 - y_0^2 - x_0^2}}\equiv  \tilde{x}_0 - 2\tilde{\epsilon}
\end{equation}
where 
\begin{equation}\label{epsilon-tilde}
	\tilde{\epsilon} = \frac{\epsilon}{\sqrt{R^2 - y_0^2 - x_0^2}} = e^{- \sigma(x_0, y_0)} \epsilon.
\end{equation}
We now use an exponential transformation $\xi(z)\equiv\exp(\I z)$ to map the flat strip into the upper half plane (UHP) and a further dilatation $\zeta(\xi)\equiv\exp(\Imag z_0) \xi$ such that the image of $\tilde{y}_0$ lies on the unitary circumference. At this point, we map the UHP into the unitary disc with a M\"obius transformation
\begin{equation}
	s = \frac{e^{\I \tilde{x}_0} - \zeta}{e^{\I\tilde{x}_0} \zeta - 1},
\end{equation}
under which the entangling point $x_0$ goes to 0 and the image of $A$ is the interval $(0,1]$ on the real line. Finally, we end up in the annulus geometry by taking the logarithm
\begin{equation}
	w \equiv \log s = \log \left [ \frac{\sin\left(\frac{z_0 - z}{2}\right)}{\sin\left(\frac{\bar{z}_0 + z}{2}\right)} \right ].
\end{equation}
Under this transformation, the boundary point of $A$ on the arctic circle $(\sqrt{R^2-y_0^2},y_0)$ is mapped to 
\begin{equation}
	w(z \equiv \I \tilde{y}_0) = \log \left [ \frac{\sin\left( \frac{\Real z_0}{2} \right)}{\sin\left( \frac{\Real z_0}{2} \right)} \right ] = 0
\end{equation}
while the boundary point on the cutoff circle $(x_0+2\epsilon,y_0)$ becomes
\begin{equation}
	w(z\equiv z_0 - 2\tilde{\epsilon}) = \log \left [ \frac{\sin \tilde{\epsilon}}{\sin\left ( \Real z_0 - \tilde{\epsilon} \right )} \right ] \approx - \log \frac{\sin \tilde{x}_0}{\tilde{\epsilon}}.
\end{equation}

Remarkably, the image $w(A)$ of the subsystem $A$ in \eqref{A} lies on the real line, i.e., $\Imag w(A)=0$. Although this is a general result at equilibrium, the same is not true for out-of-equilibrium situations, for which $w(A)$ can be a generic curve on the annulus \cite{Cardy2016}. Thanks to this simplification, the width of the annulus is simply given by
\begin{equation}\label{width}
	\widetilde{\mathcal{W}}_A =\log \frac{\sin \tilde{x}_0}{\tilde{\epsilon}}.
\end{equation}
To obtain the entanglement Hamiltonian using Eq.~\eqref{eqn:localinversetemp}, we also need the derivative of the transformation $z\to w(z)$ to the annulus. 
Therefore, for future convenience, we report
\begin{equation} \begin{split}
	w'(z) 
	      &= \frac{\sin \Real z_0}{\cos \Real z_0 - \cos(z - \I \ \Imag z_0)} = \frac{\sin \tilde{x}_0}{\cos \tilde{x}_0 - \cos(z - \I\tilde{y}_0)}.
\end{split} \end{equation}

\subsection{Entanglement entropy}\label{sec:entanglement}
Plugging Eqs.~\eqref{width} and \eqref{epsilon-tilde} into Eq. \eqref{Renyi-cft},
we get after simple algebra the R\'enyi entropy for a cutting position $x_0$  and euclidean time $y_0$ as
\begin{equation}\begin{split}\label{S-cft}
	S^{(n)}&=\frac{n+1}{12n} \widetilde{\mathcal{W}}_A = \frac{n+1}{12 n} \log \left [ \frac{e^{\sigma(x_0, y_0)}}{\epsilon(x_0,y_0)} \sin \Real z(x_0, y_0)\right ] \\[3pt]
	              &= \frac{n+1}{12 n} \log \left [ \frac{R^2 - y_0^2 - x_0^2}{\epsilon(x_0,y_0)\sqrt{R^2 - y_0^2}} \right ].
\end{split} 
\end{equation}
The UV cutoff $\epsilon$ appearing in \eqref{S-cft} is set by the inverse local fermionic density $\rho^{-1}(x_0,y_0)$ in Eq.~\eqref{dens-arctic}, 
because the latter is the only microscopic scale entering in the problem. In particular, for a connected Fermi sea one finds \cite{Dubail2017}
\begin{equation}
		\epsilon(x,y) = \frac{C_n}{\sin(\pi \rho(x,y))}	= C_n\sqrt{ \frac{R^2-y^2}{R^2-y^2-x^2}},
\end{equation}
with $C_n$ a known dimensionless non-universal constant \cite{Jin2004,Calabrese2010}. 
A more general result for split Fermi seas can be found in Refs.~\cite{Ruggiero2020,Scopa2021b,Ruggiero2021}. 
Plugging this expression in Eq.~\eqref{S-cft}, we obtain
\be\label{eqn:DWEntropy}
	S^{(n)}(x_0,y_0)= \frac{n+1}{12 n} \log \left [ \frac{(R^2 - y_0^2 - x_0^2)^{3/2}}{R^2 - y_0^2} \right ] + c_n 
\ee
$c_n \equiv -\frac{n+1}{12n} \log C_n$, and performing the analytic continuation $R\to 0$, $y_0\to \I t$
\be\label{result-Renyi}
S^{(n)}(x_0,t)= \frac{n+1}{12 n} \log \left [ t \left ( 1 + \frac{x_0^2}{t^2} \right )^{3/2} \right ] + c_n.
\ee
Finally, in the replica limit $n\to 1$, one finds the entanglement entropy
\be\label{entanglement-hydro}
S(x_0,t)= \frac{1}{6}\log\left [ t \left ( 1 + \frac{x_0^2}{t^2} \right )^{3/2} \right ] + c_1
\ee
with $c_1\simeq 0.4785$ \cite{Jin2004}, in agreement with the result of Ref.~\cite{Dubail2017} obtained with the twist field method.

\subsection{Entanglement Hamiltonian}\label{sec:ent-H-calc}
We now study the entanglement Hamiltonian, starting from the result in Eq.~\eqref{K-annulus} for the annulus and mapping it back to the arctic circle with conformal transformations. 
In particular, recalling that under conformal transformations the stress-energy tensor changes as (neglecting the Schwarzian derivative that only contributes to $\hat{K}_{A}$ with an additive constant)
\begin{equation}\label{T-transf}
	T(z) = |w^\prime(z)|^2 \; T(w(z)),
\end{equation}
we find that Eq.~\eqref{K-annulus} becomes
\begin{equation}\begin{split}
	\hat{K}_A =& \int_{\I\tilde{y}_0}^{\tilde{x}_0 - 2\tilde\epsilon + \I \tilde{y}_0} \frac{T_R(z)}{|w'(z)|} \dd z + \int_{-\I\tilde{y}_0}^{\tilde{x}_0 - 2\tilde\epsilon - \I \tilde{y}_0} \frac{{T}_L(\bar{z})}{|\bar{w}'(\bar{z})|} \dd\bar{z} \\[4pt]
	    =& \int_0^{\bar{x}_0 - 2\tilde{\epsilon}} \left | \frac{\cos\tilde{x}_0 - \cos\tilde{x}}{\sin\tilde{x}_0} \right | \left [ T_R(\tilde{x} + \I\tilde{y}_0) + {T}_L(\tilde{x} - \I\tilde{y}_0) \right ] \dd\tilde{x}.
\end{split}\end{equation}
At this point, we consider the inverse Weyl transformation $(z,\bar{z})\to(x,y)$ back to the arctic circle:
\begin{equation}
	T(x,y) = e^{-2 \sigma(x, y)} \; T(z,\bar{z})
\end{equation}
with Jacobian  $\dd\tilde{x} = e^{-\sigma(x, y)} \dd x$  and we obtain
\begin{equation}\begin{split}
	\hat{K}_A =& \int_{x_0 + 2\epsilon}^{\sqrt{R^2 - y_0^2}} \left [ \frac{T_R(x,\I y_0))}{|w'(z(x,y_0))| e^{-\sigma(x, y_0)}} +  \frac{{T}_L(x,-\I y_0)}{|\bar{w}'(\bar{z}(x,y_0))| e^{-\sigma(x, y_0)}} \right ] \dd x\\[4pt]
	    =& \int_{x_0 + 2\epsilon}^{\sqrt{R^2 - y_0^2}} \left [ (x - x_0) \sqrt{\frac{R^2 - x^2 - y_0^2}{R^2 - x_0^2 - y_0^2}} \right ] \left [ T_R(x,\I y_0) + {T}_L(x,-\I y_0) \right ] \dd x .
\end{split}\end{equation}
Finally, rotating back to real time $y_0\to\I t$ and taking the limit $R\to 0$, we find 
\begin{equation}\label{eqn:DWham}
	\hat{K}_A = \int_{x_0 + 2\epsilon}^{t} \beta(x, t) \left [ T_R(x,-t) + {T}_L(x,t) \right ] \dd x
\end{equation}
with effective inverse temperature
\begin{equation}\label{eqn:DWbeta}
	\beta(x, t) = (x - x_0) \sqrt{\frac{t^2 - x^2}{t^2 - x_0^2}}.
\end{equation}
By Taylor-expanding the effective inverse temperature~\eqref{eqn:DWbeta} around the entangling point $x_0$, 
we recover at the leading order the general prediction by Bisognano-Wichmann theorem
\begin{equation}
	\beta(x, t) \approx (x - x_0) \left [ 1 - \frac{x_0}{t^2 - x_0^2} (x - x_0) + \dots \right ] .
\end{equation}
Eq.~\eqref{eqn:DWbeta} is the major field theoretical result of this paper and fully characterise, in the scaling limit, the entanglement Hamiltonian for the domain wall melting problem. 

\section{Exact lattice results for the entanglement Hamiltonian}\label{sec:numerics}
In this section, we investigate the lattice entanglement Hamiltonian for a domain wall melting and we discuss how to correctly recover the 
field theory prediction \eqref{eqn:DWbeta}.
For the free Fermi gas, the entanglement properties (entropies and Hamiltonian) can be exactly derived from the knowledge of the two-point correlation function exploiting 
Wick's theorem. 
More precisely, the reduced density matrix $\hat\rho_A$ of an arbitrary subsystem $A$ is Gaussian~\cite{Peschel2009}
\be\label{rho-A}
\hat\rho_A\equiv\exp\left\{-2\pi\hat{K}_A\right\}=\exp \left \{ - \sum_{i, j}  \hat{c}_i^\dagger \; h_{i,j} \; \hat{c}_j  \right \}
\ee
and it is determined by the two-point correlation function $G_{i,j}\equiv \braket{\hat{c}_i^\dagger \hat{c}_j}$ restricted to $A$, i.e., $\left ( G_A \right )_{ij} = \braket{ \hat{c}_i^\dagger \hat{c}_j}_{i, j \in A}$. 
Denoting by $N$ is the number of sites in $A$, the entanglement Hamiltonian $\hat{K}_A$ is a $2^N \times 2^N$ matrix, while both $h$ and $G_A$ are $N \times N$ 
hermitian matrices. 
The latter twos can be diagonalised simultaneously and their eigenvalues are related as \cite{Peschel1999a,Chung2001,Peschel2003,Peschel2004,Peschel2009,Peschel2012}
\begin{equation}\label{relation-eig}
	e_i = \log\frac{1 - \xi_i}{\xi_i}, \quad i=1,\dots,N,
\end{equation}
where $e_i$ are the eigenvalues of $h$ and $\xi_i$ those of $G_A$. 
Eq.~\eqref{relation-eig} allows us to write the R\'enyi entropies \eqref{Renyi-entropy} as a function of the eigenvalues $\xi_i$ 
\be
S_A^{(n)}=\frac{1}{1-n}\sum_{i=1}^N \log\left[\xi_i^n +(1-\xi_i)^n\right]
\ee
and, for $n\to 1$, leads to the Von Neumann entropy
\begin{equation}\label{vN-entropy-latt}
	S_A = - \sum_{i=1}^N \left [ \xi_i \log \xi_i + (1-\xi_i) \log(1-\xi_i)\right ].
\end{equation}

\begin{figure}
\centering
\includegraphics[width=0.5\textwidth]{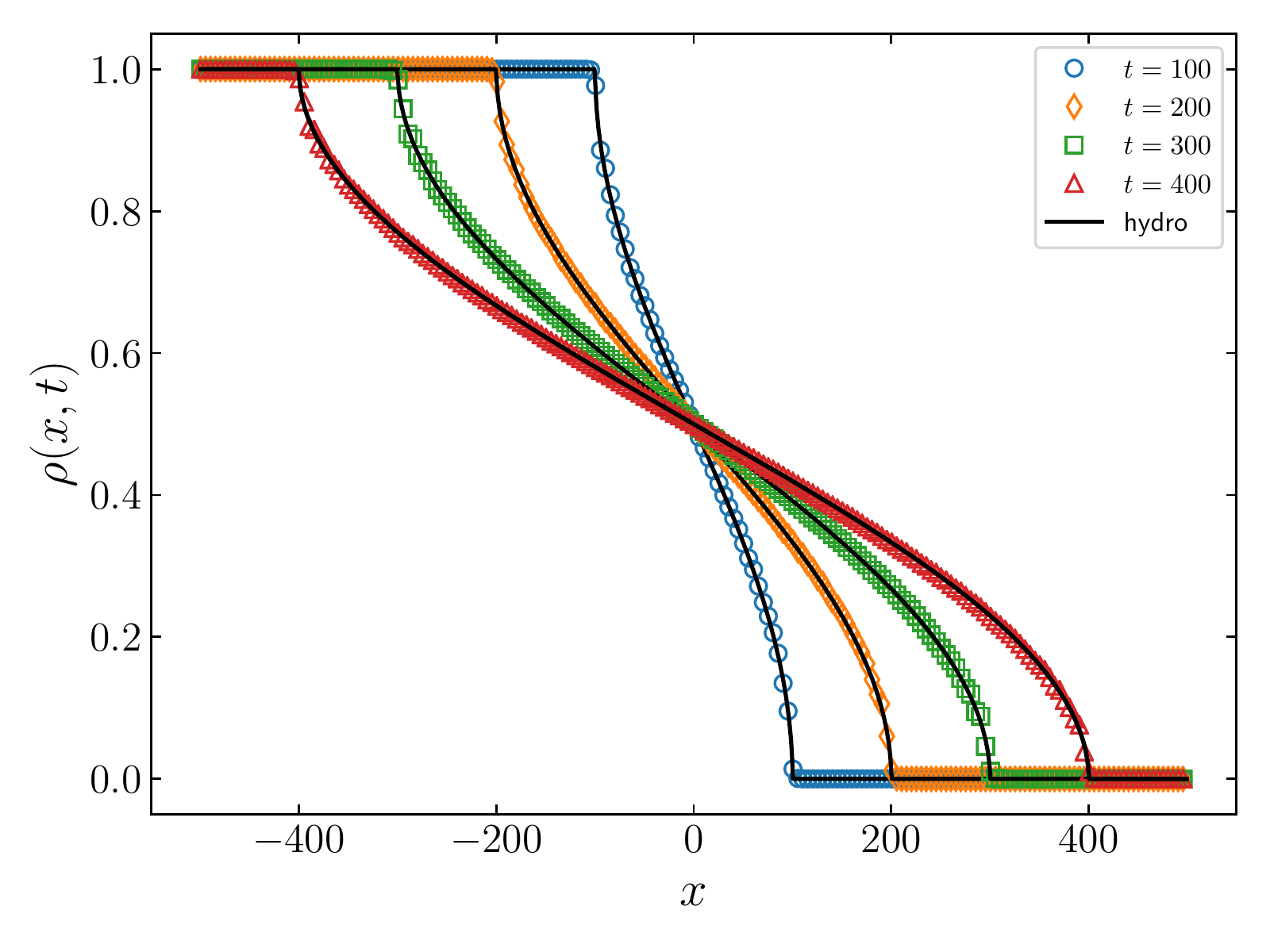}\includegraphics[width=0.5\textwidth]{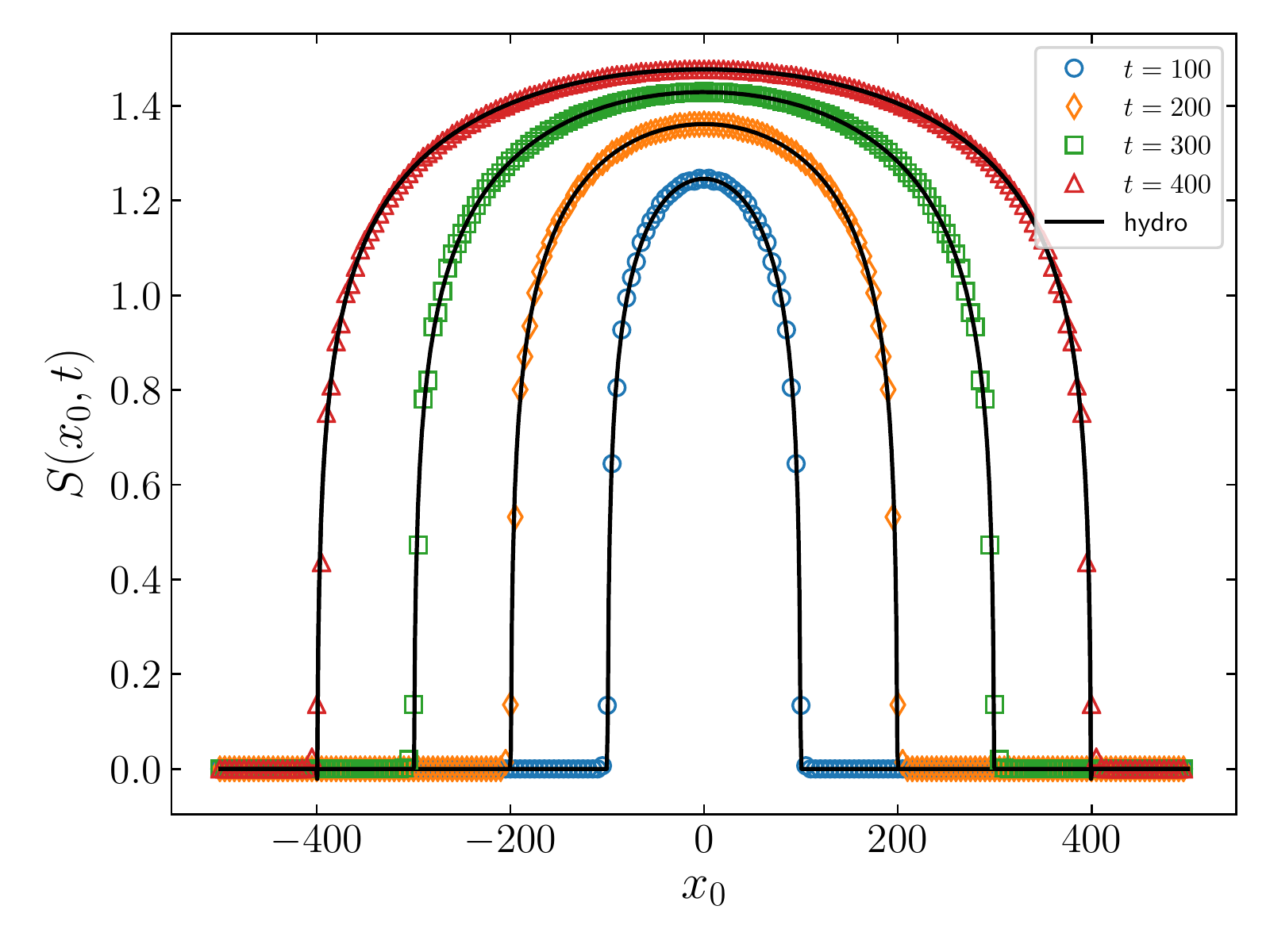}
	\caption{Snapshots of (left) the fermionic density profile and of (right) the Von Neumann  entropy after a quench from the domain wall initial state at different times. The  exact asymptotic predictions of Eqs.~\eqref{eqn:realdensity},\eqref{entanglement-hydro} ({\it black full line}) is compared with the numerical data ({\it symbols}) obtained for a system of size $L = 1000$. We observe an extremely good agreement even at short times.}\label{fig:densityentropy}
\end{figure}

Although $\hat{K}_A$ and $S_A$ are both derived from the numerical evolution of $G_A$, we notice from \eqref{vN-entropy-latt} that the larger contribution to the entanglement entropy comes from those eigenvalues $\xi_i$ of $G_A$ that significantly differ from  the values 0,1 while in the calculation of the entanglement Hamiltonian also the eigenvalues close to the edges $0$ and $1$ matter, cf. Eqs.~\eqref{rho-A}, \eqref{relation-eig}. 
As a consequence, if the numerical instabilities in the computation of $S_A$ can be easily handled introducing a cutoff for the eigenvalues, the numerical calculation of the entanglement Hamiltonian requires a high-precision algorithm, see e.g. Ref. ~\cite{DiGiulio2019,DiGiulio2020,Eisler2020,Javerzat2021}. 
In our numerical analysis, we used the open-source Python library mpmath \cite{mpmath} and we kept up to 500 digits.

As a warm up, in Fig.~\ref{fig:densityentropy}, we show the numerical results for the fermionic density $\rho(i,t)=\delta_{i,j}G_{i,j}(t)$ and for the entanglement entropy 
in the subsystem $A=[x_0,\infty]$,  together with the hydrodynamic predictions given in Eqs.~\eqref{eqn:realdensity} and \eqref{entanglement-hydro} respectively. 
The numerics are for $L=1000$ sites with $-L/2< i\leq L/2$ and the subsystem is $A=[x_0,L/2]$.  
The agreement of the hydrodynamic curves with the numerical data is extremely good.

\begin{figure}[t]
	\centering
	\includegraphics[width=0.75\linewidth]{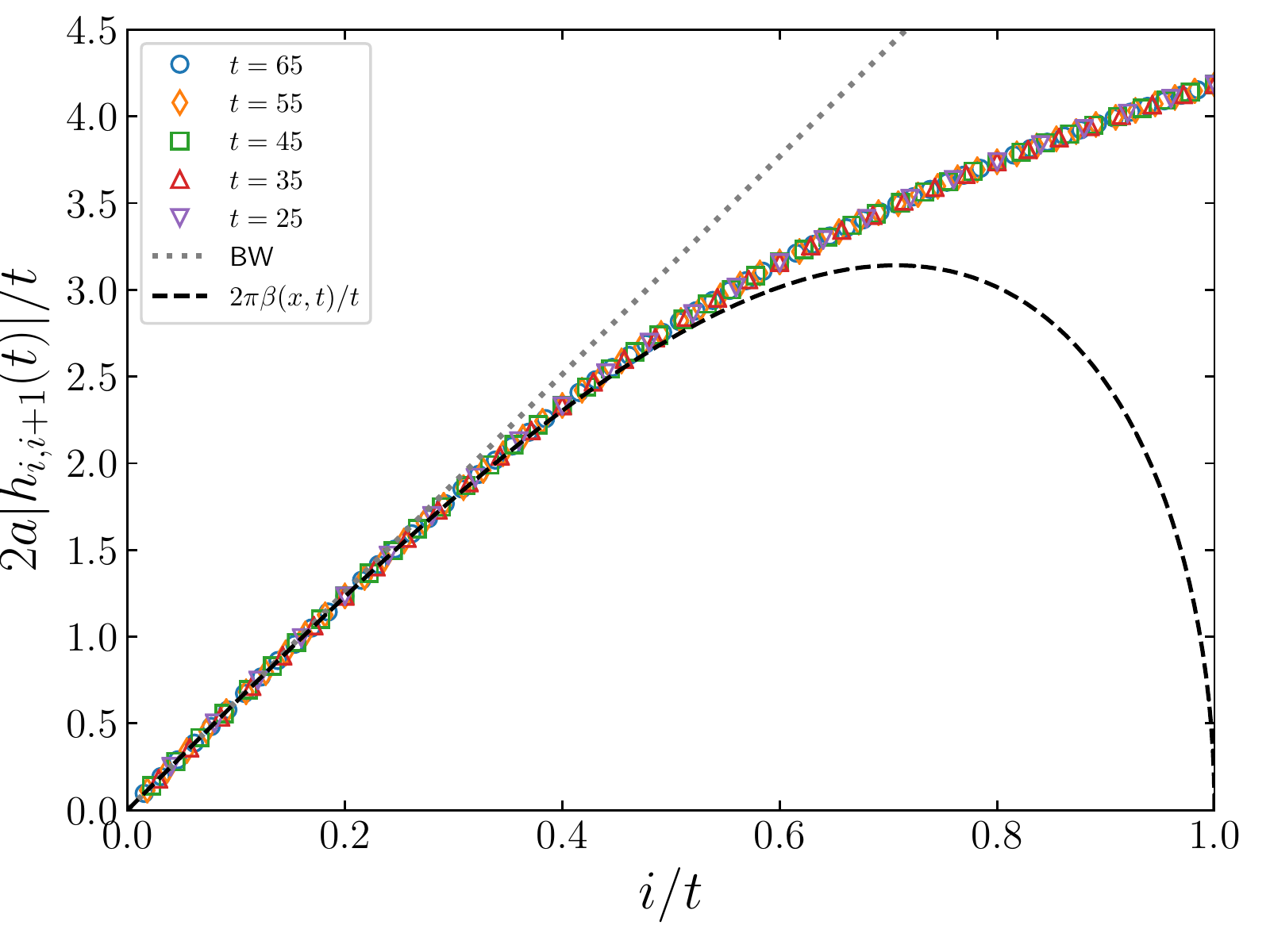}
	\caption{Nearest-neighbour entanglement Hamiltonian after the domain wall quench for the subsystem $A=[0,L/2]$, with $L=200$ and $a=1$. 
	We report the (rescaled) value of $|h_{i,i+1}|$ for different times $t$ after the quench as a function of $i/t$, showing an 
	 excellent data collapse. For $i\lesssim 0.3 t$, they match rather well the linear behaviour expected naively 
	 from the discretisation of the Bisognano-Wichmann modular Hamiltonian. 
	}\label{hiip1}
\end{figure}

\subsection{The lattice entanglement Hamiltonian}
The single-particle entanglement Hamiltonian $h_{i,j}$ in Eq.~\eqref{rho-A} is reconstructed from the eigenvalues $\xi_k$ and the eigenvectors $\phi_k(i)$ of $G_A$ as
\begin{equation}
h_{i,j}= \sum_{k=1}^N \phi_k(i) \ e_k \ \phi_k^*(j),
\end{equation}
where $e_k$ is related to $\xi_k$ by Eq. \eqref{relation-eig}. 
All the hopping elements at arbitrary distance are generically non-zero, as observed also in many physical states, see e.g. \cite{Eisler2017,ep-18,DiGiulio2019,DiGiulio2020}. 
For the domain wall melting and for the subsystem $A=[0,\infty]$, 
one would naively expect that in the hydrodynamic limit $i,t\to\infty$ with $i/t$ fixed, only the nearest neighbour term $h_{i,i+1}$ would scale like $t$ (cf. Eq.~\eqref{eqn:DWham})
while all the other hoppings would be subdominant.
To show the incorrectness of this expectation, we report the hopping elements $|h_{i,i+1}|$ in Fig.~\ref{hiip1}.
It is evident that for all $i/t$ the data collapse, but they are well reproduced by Eq.~\eqref{eqn:DWham} only for $i\lesssim 0.3 t$ (where it is actually linear and 
the correct behaviour could be inferred from the Bisognano-Wichmann theorem without performing any calculation). 
The situation is qualitatively reminiscent of what is observed in the ground state of the homogeneous system in Ref.~\cite{Eisler2017}, but there the maximal deviations 
for the hopping elements were much less pronounced (likely because of the system homogeneity).  
The solution to the problem however is the same and consists in properly taking the continuum limit of the lattice problem as outlined in Refs. \cite{Eisler2019,abch-17}
and performed in the next subsection.

\subsection{Continuum limit of the entanglement Hamiltonian}

As pointed out in Refs.~\cite{abch-17, Eisler2019}, the correct continuum limit of the entanglement Hamiltonian is achieved only retaining all long-range hoppings 
while the naive discretisation at lowest order in the derivative expansion, as seen in the previous subsection, fails. 
Hence, let us start by rewriting $\hat{K}_A$ in Eq.~\eqref{rho-A} as
\begin{equation}\label{eqn:discreteHam}
2 \pi \hat{K}_A = \sum_i \left [h_{ii} \hat{c}^\dagger_i \hat{c}_i + \sum_{r = 1}^{\infty} \left ( h_{i,i+r} \hat{c}^\dagger_{i} \hat{c}_{i+r} + \text{h.c.} \right ) \right ].
\end{equation}
For clarity, we first show how the procedure works at equilibrium (following Ref.~\cite{Eisler2019}) and after we extend this derivation to the domain wall melting.
Let us consider the continuum limit  of the lattice fermionic operators $\hat{c}_i$, $\hat{c}_i^\dagger$ in terms of chiral Dirac fermions $\psi_L$, $\psi_R$ as in Eq.~\eqref{eqn:latticefermion}. 
We obtain for the diagonal term
\begin{equation}
	h_{ii} \hat{c}^\dagger_i \hat{c}_i \sim
 a h_{x,x} \left [ \psi^\dagger_L(x) \psi_L(x) + \psi^\dagger_R(x) \psi_R(x) \right ],
\end{equation}
where we used the stationary phase approximation to cancel rapidly oscillating contributions. 
For the off-diagonal terms, using that $ \dd\varphi^\pm \approx \pm k_F \dd x \equiv \pm k_F ra$ (cf. Eq. \eqref{eqn:staticdiffphase}), one finds 
\begin{multline}
	    h_{i,i+r} \left [ \hat{c}^\dagger_{i} \hat{c}_{i+r} + \hat{c}^\dagger_{i+r} \hat{c}_{i} \right ] 
	\sim a \left ( h_{x-\frac{ra}{2},x+\frac{ra}{2}} + \frac{ra}{2} h' \right ) 2 \cos(k_Fra) \left ( \psi^\dagger_R \psi_R + \psi^\dagger_L \psi_L \right ) + \\
	     + ra^2 h_{x-\frac{ra}{2},x+\frac{ra}{2}} \left(\cos(k_Fra)  \partial_x \left ( \psi^\dagger_R \psi_R + \psi^\dagger_L \psi_L \right ) 
	     \right.\\ \left.
	   - \sin(k_Fra) \left [ \I \left ( \psi^\dagger_R \partial_x \psi_R - \psi^\dagger_L \partial_x \psi_L \right ) + \text{h.c.} \right ]\right).
\end{multline}
Integrating by parts, the terms containing $h^\prime$ are canceled and, recognising the expression for the 00 component of the stress-energy tensor  $T_{00}\equiv \I\left ( \psi_R^\dagger \partial_x \psi_R - \psi_L^\dagger \partial_x \psi_L \right )$, one obtains the continuum limit of the entanglement Hamiltonian as
\begin{equation}\begin{split}\label{K-latt-eq}
\hat{K}_A \sim& \int \frac{\dd x}{2\pi} \left [ h_{x,x} + 2 \sum_{r \geq 1} \cos(k_Fra) h_{x-\frac{ra}{2},x+\frac{ra}{2}} \right ] \left ( \psi^\dagger_R \psi_R + \psi^\dagger_L \psi_L \right ) +\\
	              &+ \int\frac{\dd x}{2\pi} \left [- 2 a \sum_{r \geq 1} r \sin(k_Fra) h_{x-\frac{ra}{2},x+\frac{ra}{2}} \right ] T_{00}.
\end{split}\end{equation}
In Eq.~\eqref{K-latt-eq}, we can identify an inverse local temperature as in Eq.~\eqref{eqn:localinversetemp}. By definition, this is given by the terms in square brackets that multiplies $T_{00}$
\begin{equation}\label{latt-beta-eq1}
	2 \pi \beta (x) \sim - 2 a \sum_{r\geq1} r \sin(k_{F} ra) h_{x-\frac{ra}{2},x+\frac{ra}{2}}
\end{equation}
while the remaining terms appearing in the first line of Eq.~\eqref{K-latt-eq} sum up to zero, see Ref.~\cite{Eisler2019} for details. 
Notice that in Eq.~\eqref{latt-beta-eq1} the sum over $r$ corresponds to sum over the anti-diagonal elements of an upper triangular matrix, due to the expansion of  $h_{x, x+ra}$ around $x + ra/2$. For the numerical implementation, it is convenient to consider a sum running over the rows of the matrix $h$ as 
\begin{equation}\label{latt-beta-eq2}
	2\pi\beta (x) \sim - 2 a \sum_{r = 1}^{r_\text{max}} r \sin(k_{F} ra) h_{i,i+r },
\end{equation}
where $r_{\rm max}$ is the maximum distance between elements that we retain in the numerics.
Differences between these two schemes \eqref{latt-beta-eq1}-\eqref{latt-beta-eq2} show up only at higher orders in the derivative expansion and thus can be neglected.

\begin{figure}[t]
	\centering
	\includegraphics[width=0.75\linewidth]{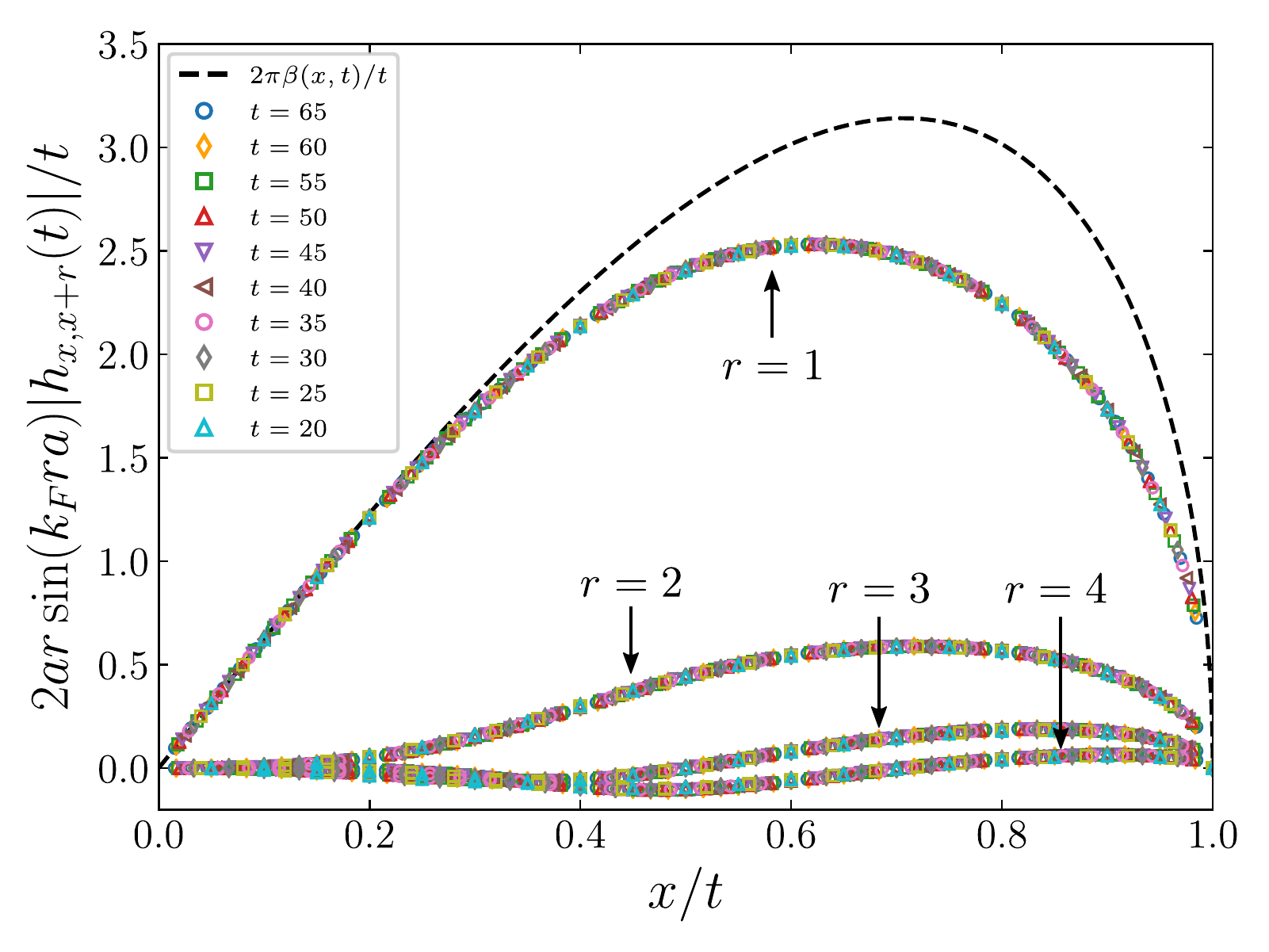}
	\caption{The elements of the sum at fixed distance $r$ entering in the final entanglement Hamiltonian \eqref{latt-beta-final}.  
	We consider the subsystem $A=[0,L/2]$ with $L=200$ and $a=1$. 
	}\label{hxxx}
\end{figure}

As anticipated, we need some little modifications of Eq.~\eqref{latt-beta-eq2} to obtain the correct lattice entanglement Hamiltonian for the domain wall melting problem. In particular, from the expression of the Fermi points $k_F^\pm$ in Eq.~\eqref{Fermi-arctic}, we obtain the phase at fixed time
\begin{equation}\label{diff-phase}
	\dd \varphi^\pm \approx k_F^\pm(x, y) r a =r\left[ \pm \arccos\Big( \frac{x}{\sqrt{R^2-y^2}}\Big) -\I \arcth\frac{y}{R}\right]
 \overset{y\to \I t; \; R\to 0}{\longrightarrow} \frac{r\pi}{2} \pm ra\pi \rho(x,t).
\end{equation}
In the last expression, we have a term proportional to the fermionic density, which is analogous to the case at equilibrium, but we do have an additional $r\pi/2$ phase. 
Using Eq.~\eqref{diff-phase} and focusing only on the terms  proportional to $T_{00}$, one finds for the off-diagonal elements of Eq.~\eqref{eqn:discreteHam}
\begin{equation}\begin{split}\label{last}
	    &h_{i,i+r}(y)  \hat{c}_i^\dagger \hat{c}_{i+r} \sim a h_{x,x+ra}(y) \left [  e^{-\I \dd\varphi^+} \psi^\dagger_R(z) \psi_R(z+ra) + e^{- \I\dd\varphi^-} \psi^\dagger_L(\bar{z}) \psi_L(\bar{z}+ra) \right ] \\
& \approx a e^{-\I r \left ( -\I \arcth(\frac{y}{R} ) \right )} h_{x,x+ra}(y)\big [- \sin(\pi \rho(x,y) ra)  \; \big [ \psi_R^\dagger \left ( \psi_R + ra\  \partial_{z} \psi_R \right ) \\[3pt]
&\qquad\qquad +  \psi_L^\dagger \left ( \psi_L + ra \ \partial_{\bar{z}} \psi_L \right ) \big ] \big ] + \text{other terms} \\
& \overset{y\to \I t; \; R\to 0}{\longrightarrow}  - ra^2 e^{-\I r \frac{\pi}{2}} h_{x,x+ra}(t) \sin(\pi\rho(x,t) ra) \big[T_L(x,t)+ {T}_R(x,-t) \big]+ \text{other terms},
\end{split}\end{equation}
where, for a better exposition,  we have not displayed those terms that goes to zero in the hydrodynamic limit $x,t\to\infty$ with $x/t$ fixed. Quite nicely, we numerically observe that the additional phase $-e^{-\I r\pi/2}$ exactly cancels the phase of $h_{i,i+r}(t)$, leading to the real quantity $|h_{i,i+r}|\equiv- e^{-\I r\pi/2} h_{i,i+r}$. Finally, from Eq.~\eqref{last} and including the hermitian conjugate, we can write the continuum limit of the entanglement Hamiltonian \eqref{eqn:discreteHam} for the domain wall quench problem as
\begin{equation}\label{KA-DW}
 \hat{K}_A \sim \int \frac{\dd x}{2\pi} \left [ 2a \sum_{r=1}^{r_\text{max}} r \sin(\pi\rho(x,t)ra) |h_{x, x+r}(t)| \right ] \left [ T _L+ T_R \right ] 
\end{equation}
from which we identify the local inverse temperature to be
\begin{equation}\label{latt-beta-final}
	2 \pi\beta(x,t) = 2a \sum_{r= 1}^{r_\text{max}} r \sin(\pi\rho(x,t)ra) |h_{x, x+r}(t)|.
\end{equation}
\begin{figure}[t]
	\centering
	\includegraphics[width=0.9\linewidth]{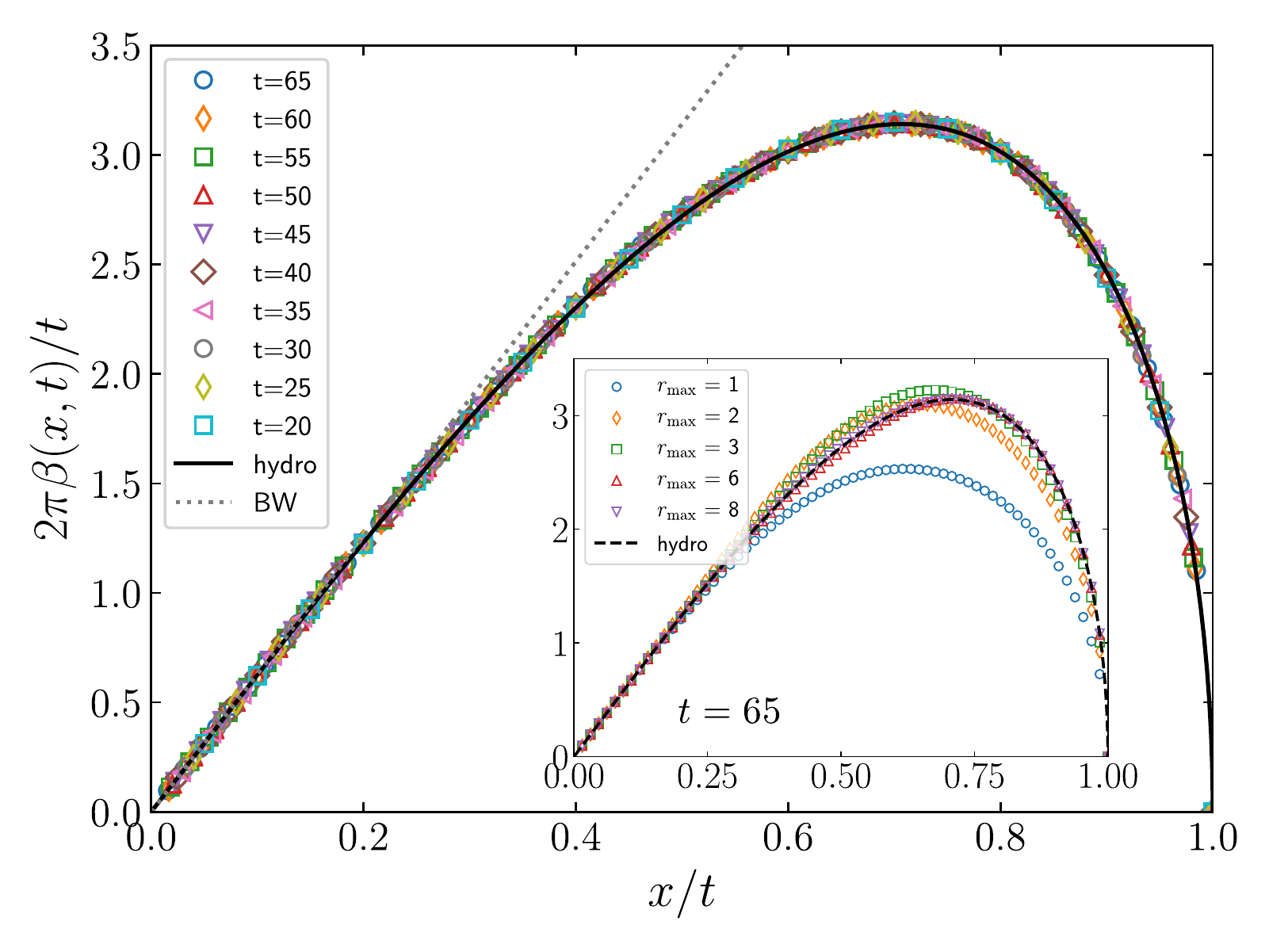}
	\caption{ Spatial profile of the inverse effective temperature $\beta(x,t)$ (rescaled with $t$) for the domain wall melting problem as a function of the scaling variable $x/t$.
	The numerical data (symbols) are obtained from Eq. \eqref{latt-beta-final} and are compared with the asymptotic prediction \eqref{eqn:DWbeta} (solid line).
	The dotted line corresponds to the linear behaviour (slope $2\pi$) from the Bisognano-Wichmann theorem.
	The data are for a system with $L=200$ sites, for $A=[0,L/2]$, with a working precision of 500 digits and with $r_{\rm max}=8$.  
	Inset: The same with different $r_{\rm max}\leq 8$.
	}\label{fig:betacollapse}
\end{figure}
Comparing this expression with the naive expectation $|h_{x,x+1}|$ (shown in Fig. \ref{hiip1}) there are two fundamental new ingredients.
First of all, because of the inhomogeneity, the Fermi momentum does depend on the position, multiplying the  naive result by $\sin(\pi\rho(x,t)ra)$.
In Fig.~\ref{hxxx} we report a few elements entering into the sum with $r\leq4$.
It is evident that the multiplication by the local Fermi momentum improves considerably the qualitative agreement of the data at $r=1$ with the 
asymptotic result. Indeed, we now observe a non-monotonic  behaviour which is forced by the fact that the density vanishes at the light cone $x=t$. 
The second ingredient is that we need to sum over the hopping terms at all distances. As evident from Fig. \ref{hxxx}, all these larger hoppings are scaling functions of 
$i/t$, but their amplitude quickly decay with $r$.  The final results of our numerical analysis for the entanglement Hamiltonian of the subsystem $A=[0,L/2]$ are reported in Fig.~\ref{fig:betacollapse}. 
We observe that by summing over the elements at various distances we obtain an extremely good agreement between numerics and 
the exact asymptotics in Eq.~\eqref{eqn:DWbeta} with $r_{\rm max}=8$ for times up to $t=65$. 
In the inset of the same figure, we report the sum truncated at different $r_{\rm max}$ for $t=65$ showing that all terms are necessary for a good match. 
For completeness,  in Fig.~\ref{fig:entH_x0} we report the same numerical analysis for $\beta(x,t)$ at different entangling point $x_0=0.25t,$ $0.5t$ ($A=[x_0,L/2]$), 
for which we observe an excellent data collapse in $x/t$ and a perfect agreement with the field theory prediction \eqref{eqn:DWbeta}.

\begin{figure}[t]
	\centering
	\includegraphics[width=0.9\textwidth]{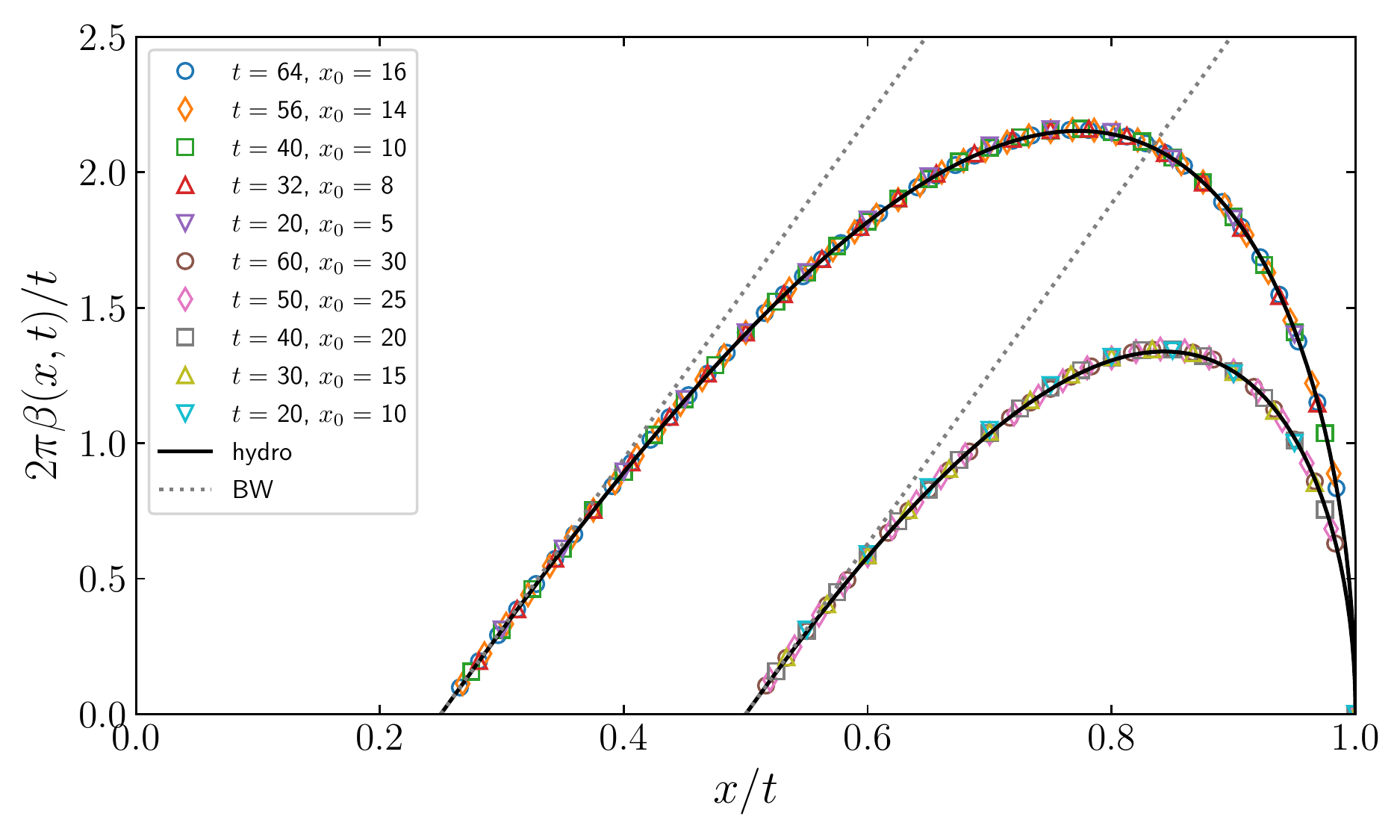}
	\caption{Spatial profile of the (rescaled) effective inverse temperature for different entangling points $x_0 / t$, as a function of the scaling variable $x / t$. 
	The outer curve is for entangling point $x_0 = 0.25 t$, while the inner has $x_0 = 0.5 t$. 
	As in Fig.~\ref{fig:betacollapse}, the numerical results (symbols) are compared with the asymptotic predictions (solid line), while the dotted line is the Bisognano-Wichmann linear behaviour. The data are for a system with $L=200$ sites and $r_{\rm max}=8$.}\label{fig:entH_x0}
\end{figure}

\section{Summary and conclusion\label{sec:conclusion}}
In this work, we considered a one-dimensional lattice gas of free fermions initially prepared in a domain wall configuration $\ket{\Psi_0}=\bigotimes_{i\leq 0} \ket{1}_i \bigotimes_{i>0} \ket{0}_i$ and subsequently let to freely expand towards the right vacuum with Hamiltonian dynamics, $\ket{\Psi(t)}=e^{-\I t \hat{H}} \ket{\Psi_0}$. For this setting, we briefly discussed the semi-classical evolution, recalling some known results about the phase-space hydrodynamics and the semi-classical profiles of conserved quantities that follow. With the goal of studying the entanglement properties of the expanding gas, we re-built quantum correlations on top of the semi-classical hydrodynamic background by expressing the latter in terms of an effective field theory for a massless Dirac fermion in a curved space-time as in Ref.~\cite{Allegra2016}. 
With this field theoretical description of the quench protocol at hand, we made use of the  annulus method \cite{Cardy2016} to obtain asymptotic predictions for the R\'enyi entropies (see Eq.~\eqref{result-Renyi} and Ref.~\cite{Dubail2017}) and for the entanglement Hamiltonian (see Eqs.~\eqref{eqn:DWham}-\eqref{eqn:DWbeta}). 
Finally, we provided high-precision numerical lattice calculations and we carefully considered the limit of large space-time scales to test our hydrodynamic result. 
We observed an excellent agreement (cf Fig.~\ref{fig:betacollapse}), already at modest system sizes and for relatively short times.

In conclusion, this work served to prove the validity of the quantum fluctuating hydrodynamics framework for the calculation of the entanglement Hamiltonian in inhomogeneous quench problems and, therefore, it opens up a wide window for future applications.  Clearly, the case of a domain wall melting  is a particularly simple instance, for which previous results for the CFT description of quantum correlations in the Luttinger regime were known \cite{Allegra2016}. To our best knowledge, a similar result is only known for the dynamics of a driven Tonks-Girardeau gas in harmonic traps \cite{Ruggiero2019}. 
A natural extension of the proposed method for generic quench settings is to join it with quantum generalised  hydrodynamics to trace backward in time the quantum correlations, similarly to what recently done for the entanglement entropies and spectrum \cite{Ruggiero2020,Collura2020,Scopa2021a,Scopa2021b,Ruggiero2021}.
An interesting application of our result would be to discretise the field theoretical result \eqref{eqn:DWham} to engineer both numerically and experimentally 
the hydrodynamic entanglement Hamiltonian of the domain wall melting, on the lines of Refs. \cite{Dalmonte:2017bzm,kbe-21,zksz-22}.

\vspace{1cm}
{\bf Acknowledgments}.  The authors acknowledge support from ERC under Consolidator grant number 771536 (NEMO). 
We acknowledge Giuseppe Di Giulio for very useful discussions. The authors are thankful to Erik Tonni and Viktor Eisler for remarks on the manuscript.

\newpage{\pagestyle{empty}\cleardoublepage}

\end{document}